\documentclass[prb,twocolumn,showpacs,amsmath,amssymb,superscriptaddress]{revtex4}
\usepackage{graphicx,ulem}
\usepackage{epstopdf}
\usepackage{enumerate}
\usepackage{epstopdf}
\newcommand \beq{\begin{eqnarray}}
\newcommand \eeq{\end{eqnarray}}

\newcommand{\bfr}{\mathbf{r}}
\newcommand{\bfk}{\mathbf{k}}

\newcommand{\bfq}{\mathbf{q}}

\newcommand{\vf}{v_F}
\newcommand{\kf}[1]{k_{F #1}}

\newcommand{\Tr}{\mathrm{Tr}}
\newcommand{\re}{\mathrm{Re}}
\newcommand{\im}{\mathrm{Im}}

\newcommand{\gp}{\hat{G}^+}

\newcommand{\ef}{\epsilon_F}
\newcommand{\matr}[4]{\left( \begin{array}{cc} #1 & #2 \\ #3 & #4 \end{array}\right)}
\newcommand{\matrd}[4]{\left( \begin{array}{cccc} #1 &  &  &  \\  & #2 &  &  \\ &  & #3 &  \\ &  & & #4 \end{array}\right)}
\newcommand{\matrx}[4]{\left( \begin{array}{cccc}  &  &  & #1 \\  & & #2 &  \\ & #3 &  &  \\ #4 &  & &  \end{array}\right)}
\begin{document}

\title{Weak Localization, Spin Relaxation, and Spin-Diffusion:\\ The  
Crossover Between Weak and Strong Rashba Coupling Limits}
\author{Yasufumi Araki}
\affiliation{
Department of Physics, University of Texas at Austin, Austin, Texas 78712, USA
}
\author{Guru Khalsa}
\affiliation{
Department of Physics, University of Texas at Austin, Austin, Texas 78712, USA
}
\affiliation{
Center for Nanoscale Science and Technology, National Institute of Standards and Technology, Gaithersburg, MD 20899, USA
}
\author{Allan H.~MacDonald}
\affiliation{
Department of Physics, University of Texas at Austin, Austin, Texas 78712, USA
}

\begin{abstract}
Disorder scattering and spin-orbit coupling are together responsible for the 
diffusion and relaxation of spin-density in time-reversal invariant systems.
We study spin-relaxation and diffusion
in a two-dimensional electron gas with Rashba spin-orbit coupling
and spin-independent disorder, 
focusing on the role of Rashba spin-orbit coupling in transport.  
Spin-orbit coupling contributes to spin relaxation, transforming the quantum interference 
contribution to conductivity from a negative weak localization (WL) correction to a positive  
weak anti-localization (WAL) correction.  The importance of spin channel mixing 
in transport is largest in the regime
% between the band-resolved and unresolved limits
where the Bloch state energy uncertainty $\hbar/\tau$ and the Rashba spin-orbit 
splitting $\Delta_\mathrm{SO}$ are comparable.  
We find that as a consequence of this spin channel mixing, the WL-WAL crossover is non-monotonic in this intermediate regime,
and use our results to address recent experimental studies of transport at two-dimensional oxide interfaces.
\end{abstract}
\maketitle

\section{Introduction}
Spin-orbit coupling, present whenever electrons move in a strong electric field,
has recently been playing a more prominent role in electronics.
%When inversion symmetry is broken spin-orbit coupling breaks the two-fold spin-degeneracy of 
%Bloch states in a crystal.  
When spin-orbit coupling is present, broken inversion symmetry lifts the two-fold spin-degeneracy of Bloch states in a crystal.
For example, as pointed out by Rashba\cite{Rashba}, spin-orbit coupling 
produces spin-splitting at surfaces and at interfaces between different materials. 
In spintronics, Rashba spin-orbit coupling can provide a handle for electrical control of spin 
since its strength and character depends not only on atomic structural asymmetry
but also on external gate voltages\cite{Koga_2002}, allowing for the possibility of a spin-based 
field-effect transistor.\cite{Datta_Das}  Alternately, spin-splitting due to Rashba spin-orbit coupling in proximity coupled  
nanowires can lead to topological superconductivity and Majorana edge states \cite{Majorana},
which can provide an attractive Hilbert space for quantum state manipulation for the purpose 
of quantum information processing.  

Because spin-orbit coupling does not conserve spin, one of its most important consequences in spintronics is its role in providing a mechanism for relaxation of non-equilibrium spin densities.  In the absence of 
spin-orbit coupling, total charge and all three components of total spin are conserved.  
%Spin relaxation mechanisms due to spin-orbit coupling can be classified into two types.
%The Elliott--Yafet (EY) mechanism,\cite{Elliott,Yafet}
%skew-scattering due to spin-orbit interactions with scattering centers
%is the the most obvious spin-relaxation process.    
%However the more subtle Dyakonov--Perel (DP) mechanism,\cite{Dyakonov-Perel}
%in which the momentum-dependent spin-orbit effective magnetic fields
%responsible for spin-splitting of the Bloch states 
%cause spin-precession between collisions, can be equally important.  
Spin relaxation mechanisms due to spin-orbit coupling can be classified into two types:
the Elliott--Yafet (EY) mechanism,\cite{Elliott,Yafet}
where skew-scattering due to spin-orbit interactions with scattering centers
is the the most obvious spin-relaxation process, and the more subtle but equally important Dyakonov--Perel (DP) mechanism,\cite{Dyakonov-Perel}
in which the momentum-dependent spin-orbit effective magnetic fields
responsible for spin-splitting of the Bloch states 
cause spin-precession between collisions.  
The DP spin relaxation mechanism is often dominant in spintronics,
and can cause subtle interplays between charge and spin transport.\cite{Burkov_2004}

\begin{figure}[tbp]
\includegraphics[width=6cm]{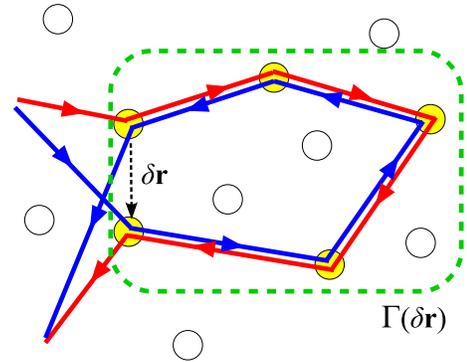}
\caption{(Color online) A schematic illustration of the the quantum correction to conductivity.
Quantum interference between a closed electron path (red) and a nearly time-reversed counterpart (blue),
alters the backscattering rate when $\bfq$, the sum of the two incoming momenta, is close to zero.   
The interference is constructive in the absence of spin-orbit coupling, enhancing back scattering and suppressing the 
diffusion constant and the conductivity, but can be destructive and enhance 
the conductivity when spin-orbit is present.}
\label{fig:paths}
\end{figure}

Spin relaxation has an important indirect effect on the quantum contribution to the conductivity.
In weakly disordered metallic systems with no spin-orbit coupling, backscattering is enhanced by constructive  
interference between time-reversed paths (see Fig.\ref{fig:paths}) yielding a negative quantum correction to the 
classical conductivity calculated from Drude's formula.  This effect is referred to as weak localization (WL).
\cite{Abrahams,Larkin,Bergmann,Montambaux}
In two-dimensional systems, WL acts as a precursor to the transition into the Anderson insulator state
in which disorder is sufficiently strong to localize electrons.
When the spin degree-of-freedom is accounted for in the absence of spin-orbit coupling, spin-degeneracy 
multiplies the conductivity correction by a factor of two.  
In general there are four two-particle spin states which contribute to the interference.
When parsed in terms of total spin eigenstates, interference between time-reversed paths
is constructive for the three triplet channels, but destructive for 
the singlet channel because of the Berry phase contributed by rotation of the spin 
wave function along the path, recovering the factor of two enhancement.
Spin relaxation changes this situation.  Because the spin-density present in the triplet channels relaxes 
their contribution to the conductivity, the correction is reduced when spin-orbit coupling is present, 
whereas the singlet channel is unaffected due to charge conservation.
When this effect is strong, either due to strong spin-orbit coupling or due to long 
phase coherence times, the quantum contribution becomes positive. 
In this case the quantum correction is referred to as weak antilocalization (WAL).
WL and WAL can be identified by studying the temperature and magnetic field dependence of 
the conductivity, since these parameters limit the phase coherence length $L$,
the characteristic length within which electrons can propagate without losing their phase coherence.
The theory of WAL onset was developed microscopically by Hikami, Larkin and Nagaoka (HLN) 
using a model with EY spin-relaxation,\cite{Hikami-Larkin-Nagaoka}
and later macroscopically using a nonlinear $\sigma$-model approach \cite{Hikami}
which demonstrated that the effect depends mainly on global symmetries and not on 
microscopic details.  Iordanskii, Lyanda-Geller and Pikus (ILP) later were the first to 
point out that DP spin relaxation also leads to WAL, and that 
the triplet channels contribution is modified compared to that implied by the EY mechanism.\cite{ILP}
WAL induced by DP spin relaxation has been identified as responsible for negative magnetoconductivity
in quantum wells \cite{Koga_2002_2,Olshanetsky_2010} and in topological insulator
surface states.\cite{Chen_2010,He_2011,Checklesky_2011}
A recent experiment on transport at the interface between
$\mathrm{LaAlO_3}$ (LAO) and $\mathrm{SrTiO_3}$ (STO) has demonstrated that 
a WL-WAL crossover can be induced by gate voltage modulation.\cite{Triscone_2010}.

Motivated partly by the experiments in LAO/STO heterostructures,
we attempt to investigate in detail the dependence of WL and WAL
transport contributions on spin-orbit coupling strength across the crossover
between resolved spin-splitting (where disorder broadening is much smaller that spin-orbit coupling) and spin-splitting obscured by disorder, by tuning the Rashba spin-orbit coupling strength in 
a two-dimensional electron gases (2DEG).
This crossover is controlled by 
a competition between two energy scales:
the Bloch state spin-splitting $\Delta_\mathrm{SO}$ induced by spin-orbit coupling in systems
without inversion symmetry, and the Bloch state energy 
uncertainty $\eta$ due to the finite lifetime of Bloch states in a disordered system.
The asymptotic behavior in the extreme cases is obvious from the arguments 
we have summarized briefly above.  In the band-unresolved limit $(\Delta_\mathrm{SO} \ll \eta)$
we can apply ILP's analysis\cite{ILP} by taking spin-orbit coupling into account perterbatively.
In this limit WAL emerges from WL behavior in the long-phase coherence limit.  
On the other hand, in the band-resolved limit $(\Delta_\mathrm{SO} \gg \eta)$, only the spin singlet channel 
contributes to quantum interference and we obtain perfect WAL behavior.
What we intend to investigate here is the behavior in the intermediate regime $(\Delta_\mathrm{SO} \approx \eta)$.
For this purpose, we have developed tools which enable us
to investigate spin-relaxation and diffusion, and to evaluate quantum corrections
to Boltzmann transport at any value of $\Delta_\mathrm{SO}/\eta$.
We have found that for a two-dimensional electron-gas model with 
Rashba spin-orbit interactions, spin relaxation in the intermediate regime cannot be simply 
described by ILP's picture.
Spin relaxation is partially suppressed by interference between channels,
leading to a new plateau on which the WL/WAL behavior is
relatively insensitive to spin-orbit coupling strength.
We suggest that such a behavior can be confirmed experimentally by tuning the spin-orbit coupling 
with gate voltage and fixing other parameters.

This paper is organized as follows.
In Section \ref{sec:theory},
we explain how we evaluate the low-energy long-wavelength limit of the 
electron-pair (Cooperon) propagator treating spin-orbit 
coupling and spin-relaxation it producees nonperturbatively.  
In Section \ref{sec:matrixelements}, we discuss our numerical results for the Cooperon
of the Rashba model, and calculate the spin relaxation lengths for each triplet channel 
in order to characterize spin relaxation behavior across the crossover between the spin-resolved and unresolved limits.
Using the spin-relaxation characteristics we have calculated, we summarize the WL-WAL crossover
in Section \ref{sec:results}, constructing a phase diagram in the $\Delta_\mathrm{SO}$-$\eta$ plane
which identifies three regimes: perfect WL, perfect WAL, and an intermediate plateau regime.
Finally, in Section \ref{sec:conclusion}, we briefly summarize our findings and present our conclusions.

\section{Microscopic theory of WL and WAL}\label{sec:theory}
As a typical example of a system with broken inversion symmetry, we consider 
an isotropic 2DEG band Hamiltonian with a Rashba spin-orbit interaction term, 
\begin{align}
\label{eq:rashba}
\hat{H}(\bfk) = \bfk^2 + \hat{\sigma}_i h_i(\bfk),
\end{align}
where we have set $\hbar=1$ so that wave vector can be identified as momentum and 
 for simplicity rescaled momentum to set $2m=1$.
We distinguish $2 \times 2$ matrices in the spin-up/down representation by a
{\it hat} accent. The second term in Eq.~\ref{eq:rashba} allows for arbitrary
spin-orbit coupling given a model with a single spin-split band.  ($\hat{\sigma}_{i} \ (i=x,y,z)$ are Pauli matrices.)
For the Rashba model, the effective magnetic field is perpendicular to momentum $\bfk$.
and has a coupling strength characterized by the parameter $\alpha$: $\mathbf{h}(\bfk) = \alpha (k_y, -k_x, 0)$.
Rashba coupling is symmetry-allowed in systems in which  
inversion symmetry is broken because the two-dimensional system is not 
a mirror plane.  For example Rashba coupling can be induced by a 
gate-induced electric-field perpendicular to the two-dimensional electron gas plane.  
It leads to spin-splitting $2\alpha k$ at momentum $k$,
where the band energies are 
\begin{align}
E_n(\bfk) = k^2+ n \alpha k,
\end{align}
with band index $n=\pm 1$.
Limiting the Fermi energy $\ef$ to be positive,
the two bands have Fermi surfaces with different Fermi radii $\kf{n}=(\vf-n \alpha)/2$,
but equal Fermi velocities $\vf=\sqrt{\alpha^2+4\ef}$.
In this article we define $\Delta_\mathrm{SO} = 2\alpha \bar{k}_F$ and 
use this number to characterize the strength of spin-orbit coupling at the Fermi energy.
Here $\bar{k}_F = (k_{F+} + k_{F-})/2 = \vf/2$ is the typical value of the Fermi momentum, independent of $n$.

We assume a disorder model with randomly-distributed, spin-independent, $\delta$-function
scatterers:
\begin{align}
\hat{H}_\mathrm{dis}(\bfr) = V \sum_{i=1}^{N} \delta(\bfr-\bfr_i),
\end{align}
where $N$ is the total number of impurities.
%Here we restrict the disorder potential isotropic both in the real space and the spin space.
After disorder averaging disorder vertices are linked in pairs 
with four-point vertex amplitude $NV^2/\Omega^2 \equiv \gamma/\Omega$,
where $\Omega$ is the volume of the system.
Thus, the disorder unaveraged one-particle Green's function
%Yasufumi - comments on disorder unaveraged Green's function
$\hat{G}_0^{\pm} = \left[\ef-\hat{H}-\hat{H}_\mathrm{dis} \pm i0\right]^{-1}$
reduces to a translationally invariant one,
\begin{align}
\hat{G}^\pm(\bfk) &= \left\langle \hat{G}_0^{\pm}(\bfk,\bfk) \right\rangle_\mathrm{dis} = \frac{1}{\ef-\hat{H}(\bfk) \pm i\eta},
\end{align}
in the Born approximation,
where $\pm$ distinguishes retarded and advanced Green's functions,
$\langle \cdot \rangle_\mathrm{dis}$ represents the average over disorder configuration,
and
\begin{align}
\eta = \frac{1}{2\tau} = -\frac{\gamma}{\Omega} \im \sum_{\bfk} G^+(\bfk) = \frac{\gamma}{4}.
\end{align}
The spectral weight of the Green's function is spread over the energy interval $\eta$,
corresponding to the finite-lifetime energy uncertainty of the Bloch states.  
When $\Delta_\mathrm{SO} \ll \eta$, the two bands are degenerate to within energy resolution and 
the role of spin-orbit interactions is simply to cause spin-precession between collisions.  
When $\Delta_\mathrm{SO} \gg \eta$, on the other hand, the two-band energies are well resolved
and coherence between bands is negligible.  In our analysis we 
assume that $\Delta_\mathrm{SO},\eta \ll \ef$, the normal experimental situation, 
but allow the ratio $\Delta_\mathrm{SO}/\eta$ to vary.  
In our discussion section, we comment briefly on the 
$\Delta_\mathrm{SO},\eta \gg \ef$ case, which corresponds closely to the 
circumstance achieved in topological insulator surface states. 

In general, the longitudinal conductivity at zero temperature is given by the Kubo--Streda formula,
\begin{align}
\sigma = \frac{1}{2\pi \Omega} \re \sum_{\bfk,\bfk'} \Tr \left\langle \hat{j}_x(\bfk) \hat{G}_0^+(\bfk,\bfk') \hat{j}_x(\bfk') \hat{G}_0^-(\bfk',\bfk) \right\rangle_\mathrm{dis},
\end{align}
where the current matrix is defined by $\hat{j}_x(\bfk) = e\hat{v}_x(\bfk) = e [\partial \hat{H}(\bfk)/\partial k_x]$.
For $\delta$-function scatterers the semi-classical Boltzmann theory result for the conductivity, namely Drude's formula, is recovered
by disorder-averaging the two Green's functions separately:
\begin{align}
\sigma_0 = \frac{1}{2\pi \Omega} \re \sum_{\bfk} \Tr \left[ \hat{j}_x(\bfk) \hat{G}^+(\bfk) \hat{j}_x(\bfk) \hat{G}^-(\bfk) \right].
\end{align}
$\sigma_0$ is proportional to the density of states at the Fermi energy.
Since the total density of states at fixed $\ef$ is independent of $\alpha$,
the classical conductivity $\sigma_0$ is independent of $\alpha$ provided that 
$\Delta_\mathrm{SO}$ is small compared to the Fermi energy $\ef$.

\begin{figure}[tbp]
\raisebox{1.5cm}{(a)} \includegraphics[width=8cm]{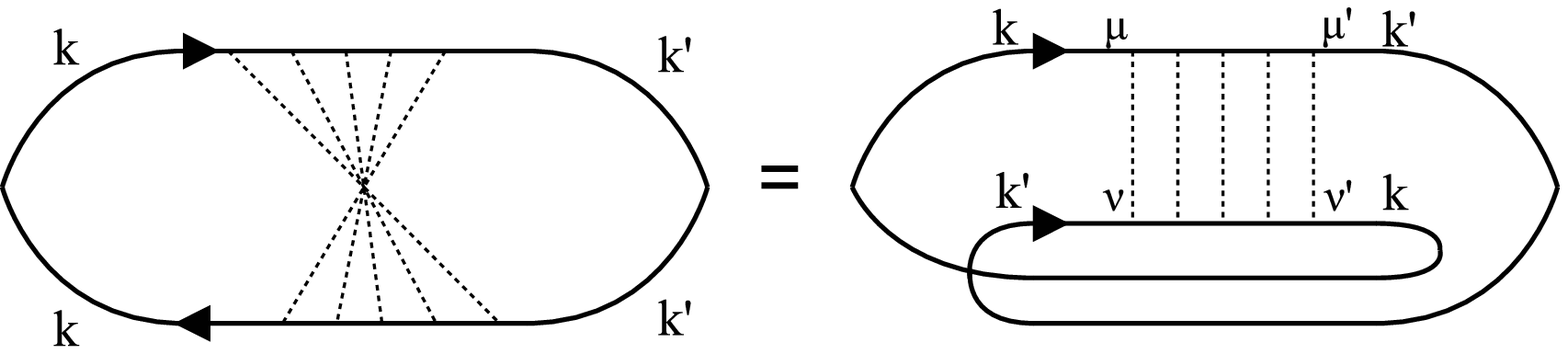}

\raisebox{2cm}{(b)} \includegraphics[width=8cm]{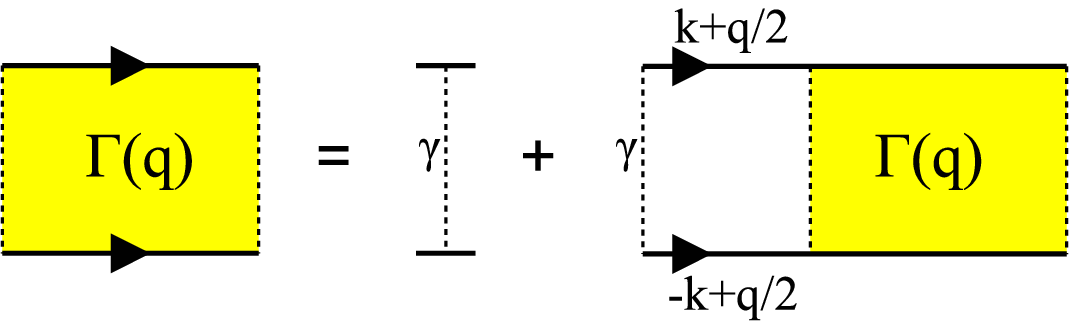}
\caption{(Color online) Feynman diagrams for the dominant quantum correction to the conductivity. (a)
The upper line and lower lines represent the retarded and advanced Green's functions $\hat{G}^\pm$, respectively.
The maximally crossed diagrams can be reorganized into a particle-particle ladder-diagram sum.  
(b) Graphical representation of the ladder diagram sum for the Cooperon which 
can be performed by solving a Bethe--Salpeter equation.}
\label{fig:cooperon}
\end{figure}

The leading quantum correction to the conductivity comes from the interference
between a closed multiple-scattering path and its time-reversed counterpart,
as illustrated in Fig.\ref{fig:paths}.
In its diagrammatic representation, the sum of this 
interference over all classical paths is captured by summing 
the diagrams in which disorder interaction lines
connecting the retarded and advanced Green's functions are maximally crossed,
as illustrated in Fig.~\ref{fig:cooperon}(a).  The particle-particle ladder diagram sum 
is referred to as the Cooperon $\check{\Gamma}(\bfq)$,
with $\bfq = \bfk + \bfk'$ the total momentum flowing into the Cooperon, or equivalently
the deviation from the perfect backscattering which occurs for $\bfq =0$.
In the following we distinguish matrices in the $4 \times 4$ tensor product space,
with the basis $\{ |\uparrow\uparrow\rangle, |\uparrow\downarrow\rangle, |\downarrow\uparrow\rangle, |\downarrow\downarrow\rangle \}$,  by a {\it check} accent over the letters as in $\check{O}$.
The contribution of the Cooperon to the conductivity is
\begin{align}
\Delta\sigma &=  \frac{\re}{2\pi\Omega} \sum_{\bfk,\bfk'} (\hat{G}^- \hat{j}_x \hat{G}^+)_{\nu' \mu}(\bfk) \check{\Gamma}^{\mu\mu'}_{\nu\nu'}(\bfq) (\hat{G}^+ \hat{j}_x \hat{G}^-)_{\mu'\nu}(\bfk') \nonumber \\
 &= \frac{e^2}{2\pi} \re \Tr \left[ \check{W} \sum_{\bfq} \check{\Gamma}(\bfq) \right], \label{eq:deltasigma}
\end{align}
where $\mu,\mu',\nu,\nu'$ take the spin indices $\uparrow$ or $\downarrow$.
We will assume that $\check{\Gamma}(\bfq)$ has a peak at backscattering $\bfq=0$, and that 
it is large only for small total momentum $\bfq=\bfk+\bfk'$. 
The area of summation by $\bfq$ is limited by the characteristic length scales of the system.
The lower cutoff is given by the inverse of a large length scale $L$, within which the electron can move without losing
its phase coherence.  $L$ acts like the (effective) size of a phase coherent system.  
The length scale $L$ decreases with increasing 
temperature due to increased inelastic scattering by phonons or other electrons,
or with an increase of magnetic field due to the cyclotron motion of the Cooper pair center of mass.
In our analysis, we represent both effects by the length $L$.
The upper wave vector cutoff is given by the inverse of the elastic mean
free path $l=\sqrt{2D\tau}$, above which electron dynamics is ballistic rather than 
diffusive.  Here $D=\vf^2\tau/2$ is the diffusion coefficient.

The {\it weight factor} ($\check{W}$) in Eq.~\ref{eq:deltasigma}
specifies how each Cooperon channel contributes to the conductivity, and is defined by
\begin{align}
\check{W}^{\mu' \mu}_{\nu' \nu} = \frac{1}{\Omega} \sum_{\bfk} (\hat{G}^- \hat{v}_x \hat{G}^+)_{\nu' \mu}(\bfk) (\hat{G}^+ \hat{v}_x \hat{G}^-)_{\mu'\nu}(-\bfk). \label{eq:weight}
\end{align}
Since the original Hamiltonian is isotropic, the matrix structure of $\check{W}$ is not 
changed by coordinate rotations which replace $\hat{v}_x$ by velocity in some other direction.

The Cooperon factor ($\check{\Gamma}(\bfq)$) in Eq.~\ref{eq:deltasigma}
is defined as an infinite sum of ladder diagrams,
\begin{align}
\check{\Gamma}(\bfq) = \frac{\gamma}{\Omega} + \frac{\gamma}{\Omega} \Omega \check{P}(\bfq) \frac{\gamma}{\Omega} + \frac{\gamma}{\Omega} \Omega \check{P}(\bfq) \frac{\gamma}{\Omega} \Omega \check{P}(\bfq) \frac{\gamma}{\Omega} + \cdots.\label{eq:cooperon}
\end{align}
The structure factor $\check{P}$ associated with a single rung of the
ladder, is given by the tensor product
\begin{align}
\check{P}(\bfq) = \frac{1}{\Omega} \sum_{\bfk} \hat{G}^+(\bfk+\tfrac{\bfq}{2}) \otimes \hat{G}^-(-\bfk+\tfrac{\bfq}{2}). \label{eq:pq}
\end{align}
Eq.(\ref{eq:cooperon}) can be summed analytically by solving an algebraic equation 
as illustrated in Fig.~\ref{fig:cooperon}(b),
to obtain 
\begin{align}
\check{\Gamma}(\bfq) = \frac{1}{\Omega} \left[\gamma^{-1} - \check{P}(\bfq) \right]^{-1}.
\end{align}
Thus the matrix structure of $\check{P}(\bfq)$ determines which Cooperon channel contribute to the conductivity correction.
Eigenvalues of $\check{P}(\bfq)$ that are close to $\gamma^{-1}$ lead to large contributions to the 
conductivity.  In the next section, we investigate the matrix structure of Cooperon in detail,
and calculate the conductivity correction as a function of the Rashba coupling constant $\alpha$
both analytically and numerically.

\section{Evaluation of the characteristic factors}\label{sec:matrixelements}
\subsection{Cooperon}
Since we expect the Cooperon to be large only in the vicinity of backscattering,
we set $\bfq = q(\cos\theta,\sin\theta)$ and expand $\check{P}(\bfq)$ in powers of $q$ up to order $O(q^2)$:
\begin{align}
\check{P}(\bfq) = \check{P}^{(0)} + q \check{P}^{(1)}_\theta + q^2 \check{P}^{(2)}_\theta +O(q^3).
\end{align}
We obtain the following expressions for the expansion coefficients:
\begin{align}
\check{P}^{(0)} &= \frac{1}{\Omega}\sum_{\bfk} \hat{G}^+ \otimes \hat{\underline{G}}^- \label{eq:p0} \\
\check{P}^{(1)}_{\theta} &= \frac{1}{2\Omega} \sum_{\bfk} \left[ (\hat{G}^+ \hat{v}_\theta \hat{G}^+) \otimes \hat{\underline{G}}^- + \gp \otimes (\hat{\underline{G}}^- \hat{\underline{v}}_\theta \hat{\underline{G}}^-) \right] \nonumber \\
\check{P}^{(2)}_{\theta} &= \frac{1}{2\Omega} \sum_{\bfk} \left[ (\hat{G}^+ \hat{v}_\theta \hat{G}^+) \otimes (\hat{\underline{G}}^- \hat{\underline{v}}_\theta \hat{\underline{G}}^-)\right], \nonumber
\end{align}
where the underlined matrices are evaluated at momentum $-\bfk$, and other matrices are 
evaluated at $\bfk$.  
$\hat{v}_\theta$ denotes the velocity projected onto the direction of 
$\bfq$: $\hat{v}_{\theta} = \hat{v}_x \cos\theta + \hat{v}_y \sin\theta$.
In the spinless (or $\alpha=0$) case the expansion simplifies to 
$P(q) = \gamma^{-1}[1-D\tau q^2]$.  
The Cooperon therefore has a pole at $q=0$ and this leads to the well-known WL correction to the 
conductivity.  Our goal here is to investigate the deviation from conventional Cooperon 
structure due to Rashba spin-orbit coupling.

The matrix structure of $\check{P}(\bfq)$ can be understood 
%by thinking about 
through the symmetries of the Hamiltonian.
Consider a spin rotation by $\pi$ around the in-plane axis perpendicular to the $\bfq$-direction generated 
by the Pauli matrix
\begin{align}
\hat{\sigma}_{\theta} \equiv \hat{\sigma}_y \cos\theta - \hat{\sigma}_x \sin\theta = \matr{0}{e^{-i(\theta-\pi/2)}}{e^{i(\theta-\pi/2)}}{0}.
\end{align}
%Its one-particle eigenstates are given by $| \pm \rangle \equiv \frac{1}{\sqrt{2}} \left[ e^{-i(\theta-\pi/2)/2} |\uparrow\rangle \pm e^{i(\theta-\pi/2)/2} |\downarrow\rangle \right]$,
%with $|\uparrow\rangle$ and $|\downarrow\rangle$ spin eigenstates in $z$-direction.
% Allan:  I reworded this.  Please check carefully.
% Yasu: -> OK.
The unitary transformation $\hat{\sigma}_\theta$ 
transforms the Rashba Hamiltonian $\hat{\sigma}_i h_i(\bfk)$ to $\hat{\sigma}_i h_i(\bfk')$,
where $\bfk'$ is the mirror reflection of $\bfk$ in a plane perpendicular to $\bfq$.
Thus, the Green's function $\hat{G}^{\pm}(\pm\bfk + \tfrac{\bfq}{2})$ gets rotated to $\hat{G}^\pm(\pm\bfk' + \tfrac{\bfq}{2})$.
By replacing the summation over $\bfk$ by one over $\bfk'$ we can conclude that  
$\check{P}(\bfq)$ in Eq.~(\ref{eq:pq}) is invariant under $\check{\Sigma}_{\theta} \equiv \hat{\sigma}_{\theta} \otimes \hat{\sigma}_{\theta}$.
It follows that $\check{P}(\bfq)$ and $\check{\Sigma}_{\theta}$ can be diagonalized simultaneously.
Since the eigenstates of $\check{\Sigma}_{\theta}$ are twofold degenerate, with the eigenvalues $\pm 1$ respectively,
$\check{P}(\bfq)$ is at least block diagonal in this basis.

Next consider spin rotation by $\pi$ around the $z$-axis which is generated by $\hat{\sigma}_z$.
Since this operation transforms the Rashba term $\hat{\sigma}_i h_i(\bfk)$ to $\hat{\sigma}_i h_i(-\bfk)$,
$\check{P}(\bfq)$ goes to $\check{P}(-\bfq)$ under the unitary transformation $\check{\Sigma}_z \equiv \hat{\sigma}_z \otimes \hat{\sigma}_z$.
Although $\check{P}(\bfq)$ is not invariant under this transformation,
even-ordered expansion terms like $\check{P}^{(0)}$ and $\check{P}^{(2)}$ are invariant.
It follows that these terms are diagonal in the representation formed by the
mutual eigenstates of $\check{\Sigma}_\theta$ and $\check{\Sigma}_z$: 
\begin{align}
\left( \begin{array}{c}
|\chi_1\rangle \\ |\chi_2\rangle \\ |\chi_3\rangle \\ |\chi_4\rangle
\end{array}\right)
=
\frac{1}{\sqrt{2}}\left(
\begin{array}{cccc}
   e^{i\theta} &   &    & -e^{-i\theta} \\
                & 1 &  1 &              \\
                & 1 & -1 &              \\
   e^{i\theta} &   &    & e^{-i\theta}
\end{array}
\right)
\left( \begin{array}{c}
|\uparrow\uparrow\rangle \\ |\uparrow\downarrow\rangle \\ |\downarrow\uparrow\rangle \\ |\downarrow\downarrow\rangle
\end{array}\right).
\label{eq:tripletsinglet}
\end{align}
This argument does not rule out off-diagonal elements in each block of $\check{P}^{(1)}$.
In fact as we emphasize later these terms do appear in $\check{P}^{(1)}$
and are responsible for anomalous 
spin relaxation behavior.
We will refer to this representation as the ``singlet-triplet basis'',
since $|\chi_3\rangle$ corresponds to the spin singlet state.
and the other three states span the three triplet states.
Note that this two-particle basis depends on the direction of $\bfq$.

\begin{figure}[tbp]
\includegraphics[width=8.5cm]{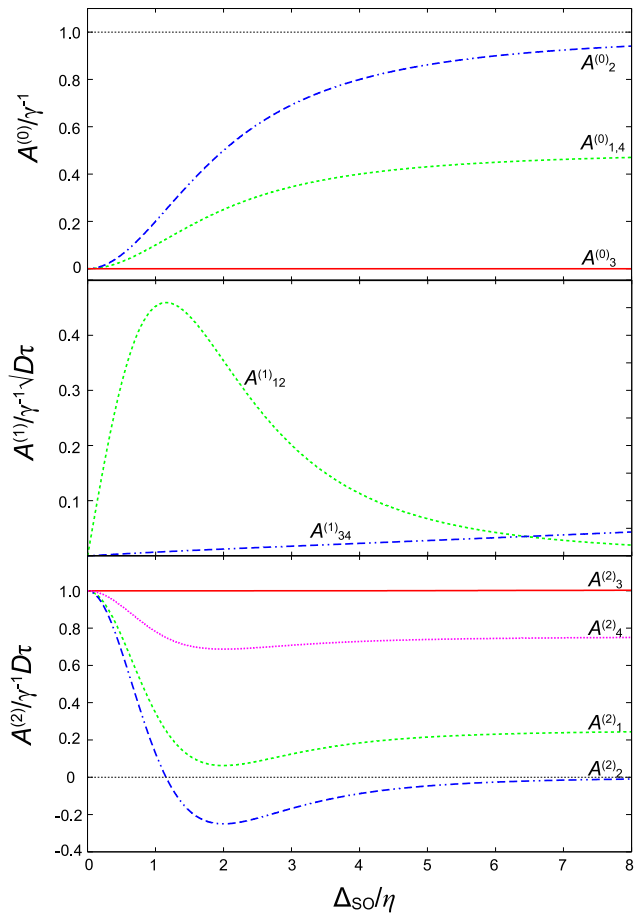}
\caption{(Color online) Matrix elements of $\check{\Gamma}^{-1}(\bfq) = \gamma^{-1}-\check{P}(\bfq)$ at zeroth,
first, and second order in a wave vector magnitude ($q$) expansion.
The illustrated calculation was for $\eta/\ef=0.01$.
The horizontal axis is the Rashba band splitting $\Delta_\mathrm{SO}=2\alpha k_F$ normalized by $\eta$.
The plotted quantities are defined in Eq.(\ref{eq:gamma-inv}).}
\label{fig:P0X}
\end{figure}

We now discuss the numerical evaluation of the $\check{P}(\bfq)$ wave-vector-magnitude 
expansion coefficients defined in Eqs.~(\ref{eq:p0})
as a function of the Rashba coupling strength $\alpha$.
We make all physical quantities dimensionless by invoking scale transformations
which reduce the Fermi energy $\ef$ and the mass $2m$ to unity.
As explained in more detail in Appendix \ref{sec:contour}, the momentum integrations can be performed analytically
by using a gradient expansion around the Fermi level and extending the integration 
contour to a closed path in the complex plane.
We calculate the structure factor matrix $\check{P}(\bfq)$ in the singlet-triplet basis motivated 
above.  The Cooperon, $\check{\Gamma}(\bfq) = \Omega^{-1} \left[\gamma^{-1} - \check{P}(\bfq)\right]^{-1}$, 
has the block-diagonal matrix structure
\begin{align}
\check{\Gamma}(\bfq) = \frac{1}{\Omega} \matr{\hat{\Gamma}_{12}(q)}{0}{0}{\hat{\Gamma}_{34}(q)},
\label{eq:gammaq}
\end{align}
with
\begin{align}
\hat{\Gamma}^{-1}_{12}(q) &= \matr{A^{(0)}_1 + q^2 A^{(2)}_1}{iq A^{(1)}_{12}}{-iq A^{(1)}_{12}}{A^{(0)}_2 + q^2 A^{(2)}_2} \label{eq:gamma-inv} \\
\hat{\Gamma}^{-1}_{34}(q) &= \matr{ q^2 A^{(2)}_3}{q A^{(1)}_{34}}{-q A^{(1)}_{34}}{A^{(0)}_4 + q^2 A^{(2)}_4}. \nonumber
\end{align}
The definitions of the coefficients $A^{(0)}$, $A^{(1)}$ and $A^{(2)}$ are given in Appendix \ref{sec:contour}.
In Fig.\ref{fig:P0X} all distinct expansion coefficients are plotted 
as a function of the Rashba coupling strength $\alpha$.
(Since $A^{(1)}$ is zero for $\alpha=0$, we normalize it by $\gamma^{-1}\sqrt{D\tau}$, 
which is comparable to $\sqrt{A^{(0)} A^{(2)}}$.)

For orientation we first comment on the characteristic behavior of
these coefficients under some extreme conditions:
\begin{itemize}
\item When the spin-orbit coupling is switched off $(\alpha=0)$,
all the constant $A^{(0)}$ and linear $A^{(1)}$ coefficients vanish,
and the quadratic coefficients $A^{(2)}$ reduce to $D\tau/\gamma$.
This result for the Cooperon leads to the 
conventional WL expression for the maximally-crossed diagram 
correction to the conductivity of a 2DEG that is free of spin-orbit coupling.  
\item When $\Delta_\mathrm{SO} \ll \eta$,
$A^{(0)}$ and $A^{(2)}$ depart from their degenerate values by $O(\alpha^2)$,
while $A^{(1)}$ is $O(\alpha^1)$.
These findings agree 
with results obtained by ILP\cite{ILP} by treating
Rashba spin-orbit coupling as a perturbation.
\item In the strong $(\Delta_\mathrm{SO} \gg \eta)$ 
spin-orbit coupling limit, $A^{(0)}$ and $A^{(2)}$ reach asymptotic values
with the ratios
\begin{align}
A^{(0)}_1 : A^{(0)}_2 : A^{(0)}_3 : A^{(0)}_4 &= 1:2:0:1 \\
A^{(2)}_1 : A^{(2)}_2 : A^{(2)}_3 : A^{(2)}_4 &= 1:0:4:3, \nonumber
\end{align}
which coincides with the behavior of the Cooperon coefficients of the massless Dirac Hamiltonian.
% Allan:  I changed this comment.  Check that you agree. 
% Yasu: ->OK. 
The Rashba and massless Dirac models agree in this limit
because the eigenstates of the two models have the same structure.
The agreement occurs even though the Rashba model normally has two Fermi surfaces,
whereas the massless Dirac model always has a single Fermi surface.  
\end{itemize}
Fig.\ref{fig:P0X} describes the crossover behavior of the Cooperon
from the weak spin-orbit coupling regime captured by ILP's analysis,\cite{ILP} 
to the strongly spin-orbit coupled limit with partial equivalence between 
massless Dirac and Rashba models.  Since the spin singlet channel is unaffected by the Rashba internal magnetic field $\mathbf{h}(\bfk)$,
the coefficients $A^{(0)}_3$ and $A^{(2)}_3$ for the singlet channel are independent of $\alpha$.

It is important here to note the behavior of the $O(q^1)$ term, which is absent in the spinless model.
Its off-diagonal components give rise to mixing between different spin channels.
Although the mixing between the singlet channel and one of the triplet channels, specified by $A^{(1)}_{34}$, 
is weak provided only that $\Delta_\mathrm{SO} \ll \ef$,
the mixing between two triplet channels, specified by $A^{(1)}_{12}$ shows quite
a nontrivial behavior.  It vanishes in both strong and weak spin-orbit coupling limits:
$A^{(1)}_{12} \propto \ef^{1/2}\Delta_\mathrm{SO}/8\eta^3$ in the band unresolved limit,
and $A^{(1)}_{12}\propto 2\ef^{1/2} \eta/\Delta_\mathrm{SO}^3$ in the resolved 
spin-splitting limit.  In the intermediate regime $(\Delta_\mathrm{SO} \approx \eta)$, on the other hand,
it is $\sim O( \sqrt{A^{(0)} A^{(2)}})$.
Due to this effect, spin relaxation is no longer described by simple exponential decay,
but rather like a damped oscillation in which spins precess as they relax.  
We elaborate on this point in the next subsection.

\subsection{Spin relaxation}

\begin{figure}[tbp]
\includegraphics[width=8.5cm]{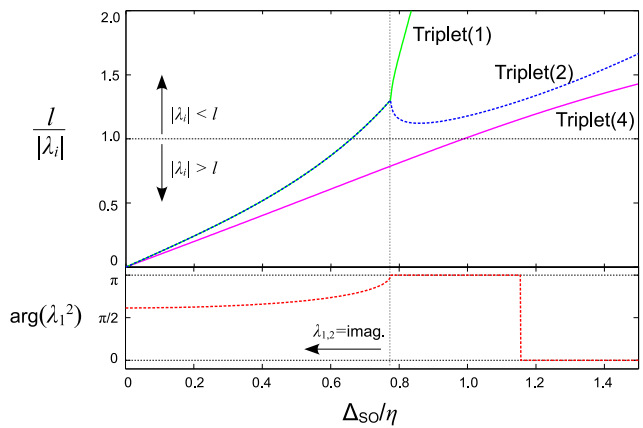}
\caption{(Color online) Upper panel: Inverse relaxation length $|\lambda_i|$ for each channel, normalized by the mean free path $l$.  The triplet channels 1,2 and 4 belong to the eigenstates of $\Gamma(q)$ in Eq. (\ref{eq:gammaq}), which are linear combinations of the triplet basis in Eq.(\ref{eq:tripletsinglet}).
Lower panel: The argument of $\lambda_1^{-2}$ normalized by $\pi$.
The values plotted in this figure were calculated for $\eta/\ef=0.01$.
It should be noted that $\lambda_{1,2}$ is complex for $\Delta_\mathrm{SO} \lesssim 0.8\eta$,
where the  $O(q^1)$ channel mixing effect is dominant.}
\label{fig:coherence-length}
\end{figure}

Using the symmetry-dictated block-diagonal structure of the $\check{P}(q)$ matrix in the singlet-triplet basis,
we can express the Cooperon in terms of its non-zero matrix elements,
\begin{align}
\check{\Gamma}(\bfq) = \frac{1}{\Omega} \left(\begin{array}{cccc}
 X_{11} & X_{12} &     0   &  0      \\
 X_{21} & X_{22} &   0      &    0    \\
  0      &    0    & X_{33} & X_{34} \\
  0     &   0     & X_{43} & X_{44}.
\end{array} \right), \label{eq:gamma'}
\end{align}
The elements $X_{ij}(q)$ are determined by inverting Eqs.(\ref{eq:gamma-inv}).
Since the singlet-triplet basis depends on the direction of the momentum $\bfq$,
we need to change the basis back to a momentum independent form
before taking the sum over $\bfq$ in Eq.~(\ref{eq:deltasigma}).
Going back to the tensor product basis and integrating out the angular dependence we obtain
\begin{align}
\sum_{\bfq} \check{\Gamma}(\bfq) = \int_{L^{-1}}^{l^{-1}} \frac{dq \ q}{2\pi} \left( \begin{array}{cccc}
 \tilde{X}_1(q) &                &                &               \\
                & \tilde{X}_2(q) & \tilde{X}_3(q) &               \\
                & \tilde{X}_3(q) & \tilde{X}_2(q) &               \\
                &                &                & \tilde{X}_1(q)
\end{array} \right), \label{eq:sum-gamma}
\end{align}
where
\begin{align}
\tilde{X}_1 = \frac{X_{11}+X_{44}}{2}, \ \tilde{X}_2 = \frac{X_{22}+X_{33}}{2}, \tilde{X}_3 = \frac{X_{22}-X_{33}}{2}.
\end{align}
Thus we need to calculate only the diagonal elements of the Cooperon matrix in Eq.(\ref{eq:gamma'}).
When there is no linear term in $q$, as in the spin-orbit decoupled limit,
each diagonal element has a diffusion peak; for example $X_{11} = [A^{(0)}_1 + q^2 A^{(2)}_1]^{-1}$.
The $O(q)$ terms in the inverse matrix  mix contributions from the two channels in each block.
The diagonal elements are then conveniently expressed in terms of partial fraction decompositions
with the form:
\begin{align}
X_{11} &= \frac{c_{11}}{\lambda_1^{-2} + q^2} + \frac{c_{12}}{\lambda_2^{-2} + q^2}, \
X_{22} = \frac{c_{21}}{\lambda_1^{-2} + q^2} + \frac{c_{22}}{\lambda_2^{-2} + q^2}, \nonumber \\
X_{33} &= \frac{c_{33}}{q^2} + \frac{c_{34}}{\lambda_4^{-2} + q^2}, \quad
X_{44} = \frac{c_{44}}{\lambda_4^{-2} + q^2} .
\end{align}
The wavelengths $\lambda$ in the denominators, whose dependence on spin-orbit coupling 
strength is plotted in Fig.~\ref{fig:coherence-length}, determine the characteristic length scales 
within which particle-hole pairs in different channels can propagate without loss.
$\lambda_{1,2,4}$ are the ``relaxation lengths'' for triplet channels,
which correspond to $\bfq$-dependent linear combinations of $|\chi_1\rangle, |\chi_2\rangle$ and $|\chi_4\rangle$. %Yasufumi
Recall that $\lambda$ is infinite in the spinless case, i.e. $\lambda_3^{-2}=0$. 

The maximally crossed diagram contribution to the conductivity is proportional to an
integral over wave vector magnitude of the diagonal elements of the Cooperon matrix.
When corrections to the conductivity are substantial this integral is dominated by contributions 
from small $q$ where our wave vector expansion is valid.  These considerations lead to a sum 
over channels of the familiar logarithmic integral: 
\begin{align}
\int_{L^{-1}}^{l^{-1}} \frac{dq \ q}{2\pi} \frac{c}{\lambda^{-2} + q^2} = \frac{c}{4\pi} \ln \frac{1 + (\lambda/l)^2}{1 + (\lambda/L)^2}. \label{eq:logl}
\end{align}
The quantum correction to the conductivity is determined by three length scales:
the mean free path $l$, the phase length $L$, which has simple power law dependences on 
temperature and magnetic field, and the spin relaxation length $\lambda$.  We note that 
\begin{itemize}
\item (i) When the spin relaxation length $\lambda$ is long compared to $L$, {\it i.e.} for $l<L \ll \lambda$,
the quantum interference is proportional to $ \ln(L/l)$, leading to its familiar
simple logarithmic temperature dependence. 
\item (ii) In the intermediate regime $l < \lambda < L$, the quantum interference correction from a particular 
channel is logarithmically dependent on its spin relaxation length $\lambda$,
{\it i.e.} it is proportional to $ \ln(\lambda/l)$.
\item (iii) When $\lambda$ is comparable to or even shorter than the mean free path $l$, {\it i.e.} for $\lambda \lesssim l$,
the quantum interference correction is absent.  Eq.(\ref{eq:logl}) approaches zero for $\lambda^{-1} \rightarrow \infty$.
\end{itemize}
As we increase the spin-orbit coupling strength $\alpha$,
the relaxation length for the spin singlet channel $\lambda_3$ remains infinite (i.e. $\lambda_3^{-2} =0$),
which implies that the logarithmic contribution from $X_{33}$ is unchanged by spin-orbit coupling.
The behavior of $\lambda_i$ for the other channels is illustrated in Fig.\ref{fig:coherence-length}.
Since $\lambda_{1,2,4}$ are comparable to the mean free path when $\Delta_\mathrm{SO}$ is 
sufficiently large compared to $\eta$,  we can see that only the spin singlet channel contributes to the quantum interference in this region.
Since we are interested in the crossover between WL and WAL behavior,
the region of $\alpha$ (or $\Delta_\mathrm{SO}$) that we need to investigate 
lies below this value.

We should note that $\lambda_{1,2}$ can take imaginary values at small $\alpha$,
because of the strong mixing between channels proportional to $qA^{(1)}_{12}$.
Up to the first order in $\alpha$, $\lambda_{1,2}^{-1}=\alpha \sqrt{(-1\pm\sqrt{7}i)/2}$.
In the context of spin diffusion, this imaginary spin relaxation implies
damped oscillations in the spin density distribution \cite{Burkov_2004}.
In our calculation, especially in case (ii) in the above classification,
it leads to a deviation from the simple logarithmic behavior $\sim \ln |\lambda/l|$,
which comes from the argument angle of $\lambda^{-1}$.
An imaginary $\lambda$ suppresses the spin channel contributions to the quantum transport 
corrections to some extent compared to the case of real $\lambda$.

\subsection{Weight factor}

The quantum transport correction contributed by each channel is also influenced by the weight 
factor factor matrix $\check{W}$, defined in Eq.~(\ref{eq:weight}).
Its evaluation is closely analogous to that required for the $O(q^2)$ term of the Cooperon.
The matrix structure in the tensor product basis is,
\begin{align}
\check{W} =
\left(
 \begin{array}{cccc}
  -R_2 &         &         & -R_1 \\
       & R_1+R_3 & -R_2'   &      \\
       & -R_2'   & R_1+R_3 &      \\
  -R_1 &         &         & -R_2
 \end{array}
\right). \label{eq:w}
\end{align}
Detailed expressions for $R_1, R_2, R_2'$ and $R_3$ 
are provided in Appendix \ref{sec:contour}.

\begin{figure}[tbp]
\includegraphics[width=8.5cm]{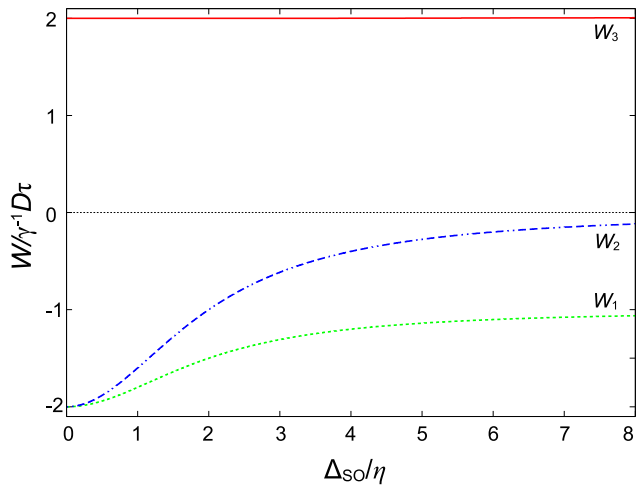}
\caption{(Color online) Behavior of the weight factors $W_{1,2,3}$ as a function of $\Delta_\mathrm{SO}/\eta$.
The results illustrated here were calculated at $\eta/\ef=0.01$.
}
\label{fig:weight}
\end{figure}

So far we have explained the matrix structure of the Cooperon and the weight factor.
Using Eqs.(\ref{eq:sum-gamma}) and (\ref{eq:w})
we find that 
\begin{align}
\Delta\sigma = \frac{e^2}{2\pi} \int_{L^{-1}}^{l^{-1}} \frac{dq \ q}{2\pi} \left[W_1 (X_{11}+X_{44}) + W_2 X_{22} + W_3 X_{33}\right], \label{eq:dels}
\end{align}
where the scalar weight factors
\begin{align}
W_1 = -R_2, \ W_2 = R_1+R_3-R_2', \  W_3 = R_1+R_3+R_2',
\end{align}
are plotted in Fig.\ref{fig:weight}.
When there is no spin-orbit coupling,
all the weight factors have the same absolute values, $W_1=W_2=-W_3=W_0$,
where $W_0 = -2D\tau/\gamma$ is the WL weight factor in the spinless case.
The minus sign in $W_3$, which is the weight factor for the spin-singlet channel, 
comes from the Berry phase due to spin rotation along the closed path.
The weight for the singlet channel, $W_3$, is essentially constant for $\alpha \ne 0$ and gives rise to WAL,
while the other weight factors which contribute to WL are suppressed.
Since the contribution from the WL channels vanishes at large spin-orbit coupling due to the spin relaxation,
only the WAL channel contributes to the quantum correction, in agreement
with conventional arguments concluding that the EY mechanism gives rise to WAL.\cite{Hikami-Larkin-Nagaoka}

\section{Total correction to the conductivity}\label{sec:results}

Finally we use Eq.(\ref{eq:dels}) to calculate the conductivity correction $\Delta\sigma$ 
as a function of the disorder amplitude $\eta$, the spin-orbit coupling strength $\alpha$,
and the phase coherence length $L$ (determined by temperature, magnetic field, etc.).
In the spinless case, we have the conventional logarithmic WL behavior, $\Delta\sigma_0(L) = -(e^2/2\pi) \ln (L/l)$.
In our calculation we calculate the ratio $r$ of the quantum correction amplitude to that of the spinless model:
\begin{align}
r(\alpha,L) = \frac{\Delta\sigma(\alpha,L)}{|\Delta\sigma_0(L)|}.
\end{align}
When there is no spin-orbit coupling, $r=-2$, {\it i.e.} two degenerate modes contribute to WL.
When the band splitting $\Delta_\mathrm{SO}$ is large enough, $r=1$, i.e. only the spin singlet mode contributes and 
it leads to WAL.  Here we calculate the detailed behavior of the ratio $r(\alpha,L)$ as $\Delta_\mathrm{SO}/\eta$ is 
varied to crossover between these two extreme limits.  

\begin{figure}[tbp]
\includegraphics[width=8.5cm]{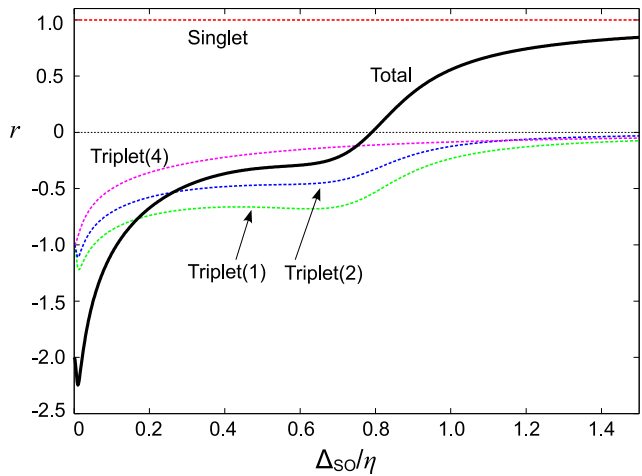}
\caption{(Color online) The quantum correction ratio $r(\alpha,L)$ as a function of the band splitting $\Delta_\mathrm{SO}=2\alpha \bar{k}_F$,
where the scattering amplitude and the phase coherence length are fixed at $\eta/\ef=0.01$ and $L/l=100$, respectively.
The dashed lines show the contributions from the singlet and the three triplet channels $X_{ii}$,
while the black bold line shows the total contribution.  The triplet channels 1,2 and 4 belong to the eigenstates of $\Gamma(q)$ in Eq. (\ref{eq:gammaq}), which are linear combinations of the triplet basis in Eq.(\ref{eq:tripletsinglet}).  Note that the WL initially strengthens, then weakens and changes to WAL.  There is a plateau in the $\Delta_{SO}$ dependence of $r$ at intermediate values.}
\label{fig:channels}
\end{figure}

\begin{figure}[tbp]
\includegraphics[width=8.5cm]{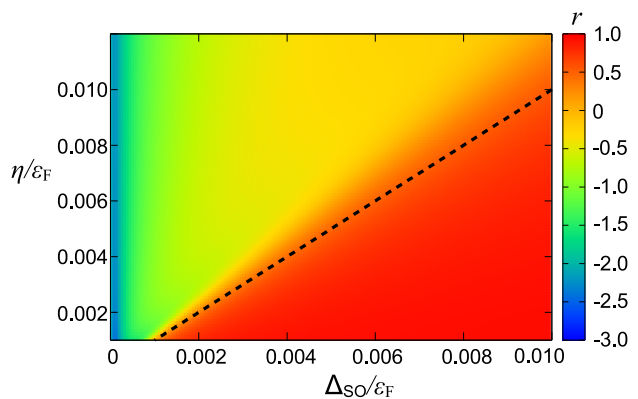}
\caption{(Color online) The quantum correction ratio $r(\alpha,L)$ {\it vs.} the band splitting 
$\Delta_\mathrm{SO}=2\alpha \bar{k}_F$
and the scattering lifetime energy-uncertainty $\eta$.
The phase coherence length is fixed at $L=2\times 10^4$, which is about 10-100 times larger than the mean free path $l$ (depending on $\eta$).
The black dashed line is $\Delta_\mathrm{SO}=\eta$.
Note that weak-localization is initially enhanced ($r < -2$) by very weak 
spin-orbit coupling.  
}
\label{fig:density1}
\end{figure}

First, we fix the scattering amplitude $\eta$ and the coherence length $L$,
and vary the Rashba coupling strength $\alpha$,
to investigate the crossover behavior going from 
full WL $(\alpha \sim 0; \; r=-2)$ to the full WAL $(\Delta_\mathrm{SO} \gg \eta; \; r=1)$.
This type of behavior is similar to that which might be expected experimentally
when gates are used to vary the Rashba coupling strength at fixed temperature.   
In Fig.\ref{fig:channels} the ratio $r(\alpha,L)$ is plotted as a function of $\alpha$,
with the other parameters fixed at $\eta/\ef=0.01$ and $L/l=100$.
We can see from this figure that an unexpected plateau-like structure appears in the intermediate region below $\Delta_\mathrm{SO} \approx \eta$,
in addition to the expected {\it perfect WAL} plateau at $\Delta_\mathrm{SO} \gg \eta$.
Interestingly this double-plateau structure cannot be described in the simple spin relaxation picture,
obtained for the extreme cases by HLN \cite{Hikami-Larkin-Nagaoka} and ILP \cite{ILP}.
% Allan:  Should repeat the references above.  
% Yasu: -> I have added the references.
To find the origin of this structure, we also plot the separate contributions from individual 
channels (i.e. $W_1 X_{11}, W_2 X_{22}, W_3 X_{33}, W_1 X_{44}$ -  see Eq.~\ref{eq:dels}).
As explained in the previous section,
Channel 3 yields a conventional logarithmic contribution to WAL,
since this channel corresponds to the spin singlet which is unaffected by spin-orbit coupling.
The other three channels give $\alpha$-dependent negative contributions.
We can see that the intermediate plateau structure comes from Channels 1 and 2,
which have imaginary relaxation lengths in the intermediate regime.
Comparing this structure with the behavior of the complex coherence lengths illustrated in Fig.\ref{fig:coherence-length},
we conclude that the crossover from the intermediate plateau to the perfect WAL plateau occurs when $\lambda$ becomes real.
This occurs around $\Delta_\mathrm{SO} \approx \eta$.  For larger values of $\Delta_{\mathrm{SO}}$
the two-channel coupling $O(q^1)$ contribution is relatively less important.
The evolution of this plateau region with $\eta$ is illustrated in Fig.\ref{fig:density1}.
Here we can clearly see that the transition between the intermediate plateau and the perfect WAL plateau occurs
around the line $\Delta_\mathrm{SO} \approx \eta$,
specified by the black dashed line in the Fig.\ref{fig:density1}.
% Allan: Is the following remark (based on your figures) consistent with ILP?
% Yasu: -> This behavior was not mentioned by ILP, because they neglected the contribution from spin channel mixing.
%          If we follow ILP's perturbative expansion more carefully, it reproduces such an enhancement of WL behavior.
%          I wonder if it shows a significant difference in experiments.
%Note that for very weak spin-orbit coupling, WL behavior is initially enhanced before the sense of the 
%change produced by spin-orbit coupling changes and the crossover to WAL begins.   
Note that for very weak spin-orbit coupling, WL behavior is initially enhanced.  The sense of the 
change produced by spin-orbit coupling then changes and the crossover to WAL begins.

\begin{figure}[tbp]
\includegraphics[width=8.5cm]{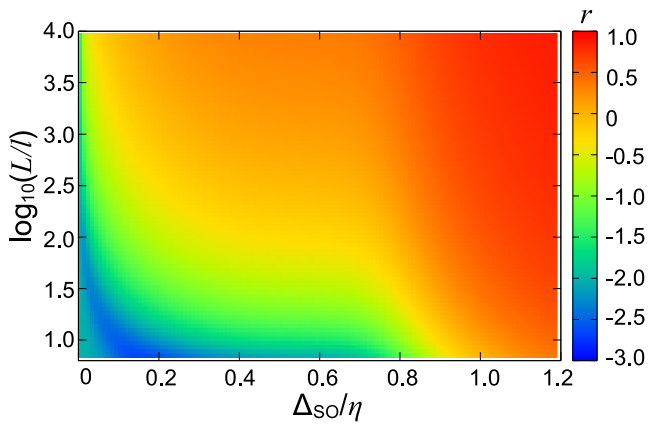}
\caption{(Color online) The quantum correction ratio $r(\alpha,L)$ as a function of the band splitting $\Delta_\mathrm{SO}=2\alpha \bar{k}_F$
and the coherence length $L$.  In this plot 
the scattering energy uncertainty is fixed at $\eta/\ef=0.01$.
All quantities are made dimensionless by setting $\ef=2m=1$.
}
\label{fig:density2}
\end{figure}

\begin{figure}[tbp]
\includegraphics[width=8.5cm]{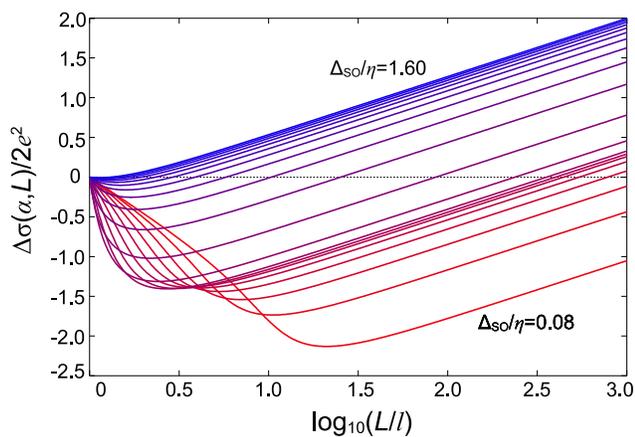}
\caption{(Color online) The behavior of the quantum correction amplitude $\Delta(\alpha,L)$ as a function of the coherence length $L$,
with the spin-orbit coupling taken at $\Delta_\mathrm{SO}/\eta = 0.08, 0.16, 0.24, \cdots, 1.60$.
The scattering amplitude is fixed at $\eta/\ef=0.01$.
All quantities are made dimensionless to setting $\ef=2m=1$.
}
\label{fig:slb}
\end{figure}

Quantum corrections to transport are normally studied experimentally 
by measuring the conductivity {\it vs.} temperature and external magnetic field.
Both temperature and external magnetic field result mainly in modulation of the phase coherence length $L$.
We therefore plot the $L$-dependence of the quantum correction ratio $r$ in Fig.\ref{fig:density2}.
In this figure we see that the weak $\Delta_\mathrm{SO}$-dependence of $r$ in the intermediate region commented on
previously appears as a crossover from WL to WAL with increasing $L$ that is more rapid than in the standard
simplified model with phenomenological triplet channel spin-lifetimes.  
The $L$-dependence of the conductivity corrections is 
plotted explicitly in Fig.\ref{fig:slb} for a series of equally spaced $\Delta_\mathrm{SO}$ values.
The intermediate and perfect WAL plateaus appear in this 
illustration as regions with densely packed lines.
The slope as a function of $L$ turns from negative (WL) to positive (WAL) around $L \sim \lambda$,
and the behavior at that length scale is unconventional for intermediate spin-orbit coupling strength.
However, we should note that it could be difficult to distinguish the intermediate plateau from the
perfect WAL plateau by performing magneto-resistance/conductance measurements,
since these measure the difference between $\Delta\sigma(L_H)$ and $\Delta\sigma(L_H=L_\epsilon)$,
where the latter is the value of $L$ in the case with no magnetic field.
Note however that the WAL differential behavior is conventional for $L > \lambda$, 
so that a relatively strong magnetic field might be necessary to observe unconventional magneto-resistance 
and this might give rise to other effects, such as classical magnetoresistance or Shubnikov--de Haas oscillation.
Since the classical conductivity is insensitive to the spin-orbit coupling strength as long as $\Delta_\mathrm{SO} \ll \ef$,
measurements of the $\alpha$-dependence at fixed temperature and magnetic field 
might be able to distinguish the two plateaus and 
might be possible if $\alpha$ is tuned by varying the electric field at 
fixed carrier density in a two-dimensional sample with both front and 
back gates.

\section{Conclusion}\label{sec:conclusion}
In this paper, we have examined the crossover behavior between WL and WAL in a two-dimensional
electron gas that is triggered by variation of the Rashba spin-orbit coupling strength.
We have used a numerical approach to evaluate the Cooperon contribution to the conductivity,
assuming only that the energy-uncertainty of Bloch states $\eta$ due to disorder scattering 
is small compared to the Fermi energy and treating spin-orbit coupling in a non-perturbative 
fashion.  
%For this reason we are able to evaluate quantum corrections to conductivity 
%for any value of the ratio of the Rashba spin-splitting   correction is 
%applicable beyond the band-unresolved to $\eta$, $\Delta_\mathrm{SO} \ll \eta$.
%When $\Delta_\mathrm{SO} \gg \eta$, there is no trace of the double degeneracy 
%present in the absence of spin-orbit coupling and each band contributes independently to the 
%conductivity.  
For this reason we are able to evaluate quantum corrections to the conductivity 
for any value of the ratio of the Rashba spin-splitting to disorder broadening $\eta$; an approach applicable beyond the band-unresolved limit when $\Delta_\mathrm{SO} \ll \eta$. 
%When $\Delta_\mathrm{SO} \gg \eta$, there is no trace of the double degeneracy 
%present in the absence of spin-orbit coupling and each band contributes independently to the 
%conductivity.  
When $\Delta_\mathrm{SO} \gg \eta$, there is no trace of the double degeneracy and each band contributes independently to the 
conductivity.  
In this limit, the system exhibits perfect WAL behavior,
where the quantum interference for spin triplet combinations quickly vanishes within the order of the mean free path.
% Allan - Please make sure that the following statement is OK? 
%although with a overall weighting coefficient which differs from that which would apply
%if the spin-orbit coupling was present only in the disorder potential.
% Yasu: -> This is not correct.
%          While the Cooperon components for the triplet channels are degenerate in HLN's calculation,
%          the degeneracy is strongly lifted in our calculation.
%          It is rather like the WAL behavior for Dirac Hamiltonian.
In the strong spin-orbit coupling limit, the Cooperon has the same structure as that for the 2D Dirac Hamiltonian,
since the band eigenstates of the two models are identical.
On the other hand, when $\Delta_\mathrm{SO} \lesssim \eta$,
a mixing between two spin triplet 
particle-hole channels at first order in a long-wavelength expansion is present,
which has been previously identified as an important feature
of spin diffusion \cite{Burkov_2004}. 
Here we show that this coupling has profound effect on quantum 
corrections to conductivity.  
%It turns the spin relaxation length is imaginary when different 
%channels are strongly coupled, and that this suppresses quantum
%corrections to conductivity.
In this regime the spin relaxation length becomes imaginary when different 
channels are strongly coupled, suppressing the damping of quantum interference
corrections to conductivity.
As a result, a new plateau-like region appears near the $\Delta_\mathrm{SO} = \eta$ line when
the maximally crossed diagrams are evaluated as a function of spin-orbit coupling strength
at fixed phase coherence length.
%Yasufumi - I removed "(" here.
Although it seems difficult to identify these two plateaus by the investigation of the differential behavior by $L$,
like the magneto-resistance/conductance measurement under a finite magnetic field,
we suggest that they can be distinguished by tuning the spin-orbit coupling strength by an external gate voltage
and fixing the coherence length (temperature and magnetic field).

Although we have limited our attention here to a simple model with spin-independent 
disorder scattering and a single spin-split band that has  
circularly symmetric Fermi surfaces, the numerical approach we have taken is 
readily generalized to an arbitrary band model and to models with spin-dependent disorder
scattering.  
% Allan:  This statement OK? -> Yasu: OK.
Dealing with anisotropy requires only that an angular average over the Fermi surface 
be added to sums over band state labels. 
% Allan:  Following discussion OK? -> Yasu: OK.
Qualitative aspects of the Rashba model results reported on here apply to
other two-dimensional electron systems with broken inversion symmetry.
For two-dimensional electron systems, inversion symmetry can usually
be varied {\it in situ} by tuning gate voltages.  
For any two-dimensional electron system without inversion symmetry, the double spin-degeneracy of the Bloch bands 
is lifted by spin-orbit coupling.  When the spin-splitting $\Delta_{\mathrm SO}$ is larger than 
the Bloch state energy uncertainty $\eta$, the spin-split bands can be viewed as 
distinct independent bands with momentum-dependent spin-orientations.  
It follows that in this regime, the spin-relaxation length is on the order the mean-free path,
{\it i.e.} spin-memory is lost at every collision.  Once this occurs we do not expect to 
see a crossover from WAL to WL when the phase length $L$ is decreased by 
increasing the magnetic field or decreasing temperature.  At weaker spin-orbit coupling 
strengths we do expect to see WL at some temperatures and fields.
However, our study shows that the way in which a WL regime emerges at 
weaker spin-orbit coupling can be nontrivial and is determined by specific features 
of the band structure of a particular system.  

One potentially interesting application of our approach is to two-dimensional electron gases formed at 
oxide heterojunctions, for instance to the $t_{2g}$ electron-gas systems at the interface between
LaAlO$_3$ and SrTiO$_3$.  The presence in this case of three different orbitals,
each of which can have Fermi surfaces, might make the spin relaxation scenario rich \cite{LAO-STO}.
% Allan:  Appropriate references below?  Comment OK? -> Yasu: I have added some references Guru has mentioned.
It is known that Rashba spin-orbit splitting in these systems \cite{Guru,Zhong}
is strongest near the avoided crossing between two higher energy (lower density) elliptical $xz,yz$ subbands
and a lower energy (higher density) $xy$ subband.  There are indeed indications 
that magnetoresistive transport anomalies occur when the Fermi level is near these 
weakly avoided crossings \cite{Joshua}.
Another potentially interesting system is two dimensional electron gases formed 
in the layers of transition metal dichalcogenide two-dimensional materials.
Spin-orbit coupling and band spin-splitting is particularly strong in the valence bands 
of this class of materials.
Coupling between spin and other degrees of freedom $\mathrm{MoS_2}$ \cite{DiXiao_2012},
may give rise to interesting complex behavior \cite{DiXiao_2013}, although 
we note that studies of transport in these materials are at a very early stage \cite{Neal_2013}.
% Allan:  Are there any experimental papers on weak-localization in TMD's? -> Yasu: Ref.27 is one of those.

\begin{acknowledgments}
YA is supported by a Japan Society for the Promotion of Science Postdoctoral Fellowship for Research Abroad (No.25-56).
Work at the University of Texas was supported by the Welch Foundation under grant
TBF1473 and by the DOE under Division of Materials Science and Engineering 
grant DE-FG03-02ER45958.
\end{acknowledgments}

\

\appendix

\section{Calculating the Cooperon and the weight factor by contour integration} \label{sec:contour}

In this section, we discuss in detail the procedure we use to obtain Cooperon and weight factor matrices.
The key ingredient here is to split the Green's function into a sum of contributions from each band:
\begin{align}
\hat{G}^\pm(\bfk) &= \sum_n \frac{\psi_n(\bfk) \psi^\dag_n(\bfk)}{g^\pm_n(\bfk)},
\end{align}
where $\psi_n(\bfk)$ is the eigenfunction for band $n(= \pm 1)$ in momentum and spin representation,
and $g^\pm_n(\bfk) = \ef-E_n(\bfk)\pm i\eta$.
In our 2DEG model, the Green's function simplifies to,
\begin{align}
\hat{G}^\pm(\bfk) = \sum_n \frac{1}{2g^\pm_n(k)} \matr{1}{-in e^{-i\phi}}{in e^{i\phi}}{1}.
\end{align}
Since the band structure is isotropic, the denominator is independent of $\phi$,
the direction of the wave vector $\bfk$.
Similarly, the velocity matrix $\hat{v}_\theta = \hat{v}_x\cos\theta +\hat{v}_y\sin\theta$ can be written as 
\begin{align}
\hat{v}_\theta(\bfk) = \sum_n \frac{v_n(k)}{2} \matr{\cos(\theta-\phi)}{in e^{-i\theta}}{-in e^{i\theta}}{\cos(\theta-\phi)}.
\end{align}

Using these expressions, the sum over $\bfk$ in Eq.(\ref{eq:p0}) can be separated 
into integrations over the orientation $\phi$ and the modulus $k$ of band wave vector.
The phase integration eliminates elements which vary like $\exp(im\phi)$ for some non-zero value 
of $m$ and hence determines the matrix structure.
The integrations over $k$ have the general form
\begin{align}
\int_0^{\infty} dk \ k \frac{f^{n_1 \cdots n_j}_{m_1 \cdots m_j}(k)}{g^+_{n_1}(k) \cdots g^+_{n_j}(k) \ g^-_{m_1}(k) \cdots g^-_{m_j}(k)}.
\end{align}
They can be completed by extending the path of integration into a large contour in the complex plane and
using Cauchy's theorem.  If we choose to close the contour in the upper half complex plane,
the poles $\bar{k}_n$ are given by the solutions of the equations $g^+_n(\bar{k}_n)=0$.
As long as we limit the disorder strength to lie within the diffusive regime $\eta \ll \ef$,
we can solve this equation by using a gradient expansion around the Fermi surface,
\begin{align}
g^+_n(\bar{k}_n) \simeq v_{F}(\bar{k}_n-k_{Fn})+i\eta.
\end{align}
By summing over the band indices, we determine the values of the matrix elements.
Because the resulting expressions are extremely cumbersome, we have evaluated 
the residues and summed over band indices numerically.
In the following subsections, we show how the matrix elements can be constructed at each order in $q$-expansion.

\subsection{$O(q^0)$}
To leading order in the $q$-expansion,
we obtain the form
\begin{align}
\check{P}^{(0)} =  p^{(0)}_1 \matrd{1}{1}{1}{1} - p^{(0)}_2 \matrx{0}{1}{1}{0}
\end{align}
in the tensor product basis,
where
\begin{align}
p^{(0)}_1 &= \sum_{n,m} I^{(0)}_{n,m}, \quad p^{(0)}_2 = \sum_{n,m} I^{(0)}_{n,m} nm, \\
I^{(0)}_{n,m} &= \frac{1}{2\pi} \int dk \frac{k}{4g_n^+ g_m^-}.
\end{align}
The coefficients in the singlet-triplet basis are
\begin{align}
A^{(0)}_1 = A^{(0)}_4 = \gamma^{-1} - p^{(0)}_1, \ A^{(0)}_2 = \gamma^{-1} - p^{(0)}_1 + p^{(0)}_2.
\end{align}
We can show analytically that $A^{(0)}_3 = \gamma^{-1} - p^{(0)}_1 - p^{(0)}_2$ vanishes at any value of $\alpha$:
Since $p^{(0)}_1 + p^{(0)}_2 = \sum_{n,m}(1+nm)I^{(0)}_{n,m}$ vanishes when $nm=-1$,
only particle-hole pairs with band indices $n=m$ contribute to $A^{(0)}_3$.
Taking the residual value, we obtain
\begin{align}
I^{(0)}_{n,n} = \frac{i \bar{k}_n}{-4\vf g^-_n(\bar{k}_n)} \sim \frac{k_{Fn}}{8\vf\eta} = \frac{1}{4\gamma},
\end{align}
where we have used $g^-_n(\bar{k}_n) = g^+_n(\bar{k}_n)-2i\eta = -2i\eta$.
Therefore, $p^{(0)}_1 + p^{(0)}_2 = \gamma^{-1}$,
which leads to $A^{(0)}_3=0$.

\subsection{$O(q^1)$}
At linear and the quadratic orders in the $q$-expansion, we should note that $\hat{G}^\pm \hat{v}_\theta \hat{G}^\pm$ in Eq.(\ref{eq:p0})
can be decomposed as
\begin{align}
& (\hat{G}^{\pm}\hat{v}_\theta \hat{G}^{\pm})(\bfk) = \sum_{n_1 n_2 n_3} \frac{v_{n_2}}{8g^{\pm}_{n_1} g^{\pm}_{n_3}}\left(n_2 -\frac{n_1+n_3}{2}\right) \nonumber \\
& \quad \times \Biggl[ \matr{0}{ie^{-i\theta}}{-ie^{i\theta}}{0} -(n_1+n_3)\cos(\theta-\phi)\matr{1}{0}{0}{1} \nonumber \\
& \quad \quad \quad +n_1 n_3\matr{0}{i e^{i(\theta-2\phi)}}{-i e^{i(2\phi-\theta)}}{0} \Biggr]. \label{eq:gvg}
\end{align}
Substituting this decomposition to Eq.(\ref{eq:p0}) and integrating out the phase $\phi$,
we obtain the matrix decomposition of $\check{P}^{(1)}_\theta$ in the tensor product basis,
\begin{align}
& \check{P}^{(1)}_\theta = \\
& p^{(1)}
\left(
 \begin{array}{cccc}
                &               & \Theta^* &               \\
                &               &               & \Theta^* \\
  \Theta &               &               &               \\
                & \Theta &               &
 \end{array}
\right)
- (p^{(1)})^*
\left(
 \begin{array}{cccc}
               & -\Theta^* &                &               \\
  -\Theta &                &                &               \\
               &                &                & \Theta \\
               &                & \Theta^*  &              
 \end{array}
\right), \nonumber
\end{align}
with the shorthand notation $\Theta = -ie^{i\theta}$.
Here the factor $p^{(1)}$ is defined by
\begin{align}
& p^{(1)} = \frac{1}{2} \sum_{n m_1 m_2 m_3} \left[ \left( I^{(1)}_{n m_1 m_2 m_3} \right)^* - n m_1 I^{(1)}_{n m_1 m_2 m_3} \right], \\
& I^{(1)}_{n m_1 m_2 m_3} = \frac{1}{2\pi} \int dk \frac{k v_{m_2}}{16 g^+_{n} g^-_{m_1} g^-_{m_3}} \left(m_2 -\frac{m_1+m_3}{2}\right).
\end{align}
Applying the unitary transformation by $\check{T}_\theta$, we obtain the correspondence to the coefficients in the singlet-triplet basis,
\begin{align}
A^{(1)}_{12} = \re\ p^{(1)}, \quad A^{(1)}_{34} = -\im\ p^{(1)}.
\end{align}
It should be noted that the linear term in $q$ is not diagonal even in the singlet-triplet basis,
and accounts for the coupling between different channels at finite momentum.

\subsection{$O(q^2)$}
Substituting the decomposition in Eq.(\ref{eq:gvg}) to Eq.(\ref{eq:p0}),
we obtain the matrix decomposition of $\check{P}^{(2)}_\theta$ in the tensor product basis,
\begin{align}
\check{P}^{(2)}_\theta &= -p^{(2)}_1 \matrx{-e^{-2i\theta}}{1}{1}{-e^{2i\theta}} + p^{(2)}_2 \matrd{1}{1}{1}{1} \nonumber \\
 & \quad \quad - p^{(2)}_3 \matrx{0}{1}{1}{0},
\end{align}
with the coefficients
\begin{align}
& p^{(2)}_1 = \sum_{\{n,m\}} I^{(2)}_{\{n,m\}}, \quad p^{(2)}_2 = \sum_{\{n,m\}} 2n_1 m_1 I^{(2)}_{\{n,m\}}, \\
& p^{(2)}_3 = \sum_{\{n,m\}} n_1 n_3 m_1 m_3 I^{(2)}_{\{n,m\}}, \nonumber \\
& I^{(2)}_{\{n,m\}} = \int \frac{dk}{2\pi} \frac{k v_{n_2} v_{m_2} (2n_2 -n_1 -n_3)(2m_2 -m_1 -m_3)}{512 g^+_{n_1} g^+_{n_3} g^-_{m_1} g^-_{m_3}},
\end{align}
where $\{n,m\} = \{ n_1,n_2,n_3,m_1,m_2,m_3 \}$.
This can be diagonalized by the unitary transformation $\check{T}_\theta$,
which leads to the following connection to the singlet-triplet basis:
\begin{align}
A^{(2)}_1 = -p^{(2)}_1+p^{(2)}_2, & \quad A^{(2)}_2 = -p^{(2)}_1+p^{(2)}_2-p^{(2)}_3, \\
A^{(2)}_4 = p^{(2)}_1+p^{(2)}_2, & \quad A^{(2)}_3 = p^{(2)}_1+p^{(2)}_2-p^{(2)}_3. \nonumber
\end{align}

\subsection{Weight factor}
The decomposition in Eq.(\ref{eq:gvg}) can also be applied to the calculation of the weight factor matrix.
Substituting the decomposition to the definition of weight factor matrix in Eq.(\ref{eq:weight}) and integrating out the phase $\phi$,
we obtain the form in Eq.(\ref{eq:w}), with
\begin{align}
& R_1 = \sum_{\{n,m\}} J_{\{n,m\}}, \quad R_2 = \sum_{\{n,m\}} (n_1 m_3 + n_3 m_1)J_{\{n,m\}}, \\
& R'_2 = \sum_{\{n,m\}} (n_1 m_1 + n_3 m_3)J_{\{n,m\}},\nonumber \\
& R_3 = \sum_{\{n,m\}} n_1 n_3 m_1 m_3 J_{\{n,m\}}, \nonumber \\
& J_{\{n,m\}} = \nonumber \\
& \int \frac{dk}{2\pi} \frac{k v_{n_2} v_{m_2}}{64 g_{m_1}^+ g_{n_3}^+ g_{n_1}^- g_{m_3}^-} \left(n_2-\tfrac{n_1+n_3}{2}\right) \left(m_2-\tfrac{m_1+m_3}{2}\right). \nonumber
\end{align}
The definition of $J_{\{n,m\}}$ looks similar to $I^{(2)}_{\{n,m\}}$,
while the difference appears in the retarded/advanced indices in the denominator.
We should note that $(n_1,n_3)$ and $(m_1,m_3)$ cannot be exchanged here, respectively.

%%% REFERENCES %%%


\begin{thebibliography}{99}

\bibitem{Rashba}
Yu.~A.~Bychkov and E.~I.~Rashba,
J.~Phys.~C \textbf{17}, 6093 (1984).

\bibitem{Koga_2002}
J.~Nitta, T.~Akazaki, H.~Takayanagi, and T.~Enoki,
Phys.~Rev.~Lett.~\textbf{78}, 1335 (1997);
% ``Gate Control of Spin-Orbit Interaction in an Inverted In0.53Ga0.47As/In0.52Al0.48As Heterostructure''
% http://dx.doi.org/10.1103/PhysRevLett.78.1335
T.~Koga, J.~Nitta, H.~Takayanagi, and S.~Datta,
Phys.~Rev.~Lett.~\textbf{88}, 126601 (2002).
% ``Spin-Filter Device Based on the Rashba Effect Using a Nonmagnetic Resonant Tunneling Diode''
% http://dx.doi.org/10.1103/PhysRevLett.88.126601

\bibitem{Datta_Das}
S.~Datta and B.~Das,
Appl.~Phys.~Lett.~\textbf{56}, 665 (1990).
% ``Electronic analog of the electro‐optic modulator''
% http://dx.doi.org/10.1063/1.102730

\bibitem{Majorana}
J.~Alicea,
Rep.~Prog.~Phys.~\textbf{75}, 076501 (2012).
%New directions in the pursuit of Majorana
%fermions in solid state systems

\bibitem{Elliott}
R.~J.~Elliott,
Phys.~Rev.~\textbf{96}, 266 (1954).
% ``Theory of the Effect of Spin-Orbit Coupling on Magnetic Resonance in Some Semiconductors''
% http://dx.doi.org/10.1103/PhysRev.96.266

\bibitem{Yafet}
Y.~Yafet,
Solid State Phys.~\textbf{14}, 1 (1963).
% ``g Factors and Spin-Lattice Relaxation of Conduction Electrons''
% http://dx.doi.org/10.1016/S0081-1947(08)60259-3

\bibitem{Dyakonov-Perel}
M.~I.~D'yakonov and V.~I.~Perel',
Sov.~Phys.~Solid State \textbf{13}, 3023 (1972).

\bibitem{Burkov_2004}
A.~A.~Burkov, A.~S.~N\'{u}\~{n}ez, and A.~H.~MacDonald,
Phys.~Rev.~B \textbf{70}, 155308 (2004).
% ``Theory of spin-charge-coupled transport in a two-dimensional electron gas with Rashba spin-orbit interactions''
% http://dx.doi.org/10.1103/PhysRevB.70.155308

\bibitem{Abrahams}
E.~Abrahams, P.~W.~Anderson, D.~C.~Licciardello, and T.~V.~Ramakrishnan,
Phys.~Rev.~Lett.~\textbf{42}, 673 (1979).
% ``Scaling Theory of Localization: Absence of Quantum Diffusion in Two Dimensions''
% http://dx.doi.org/10.1103/PhysRevLett.42.673

\bibitem{Larkin}
L.~G.~Gorkov, A.~I.~Larkin, and D.~E.~Khmel'nitzkii,
JETP Lett.~\textbf{30}, 228 (1979);
% ``Particle conductivity in a two-dimensional random potential''
% http://www.jetpletters.ac.ru/ps/1364/article_20629.shtml
B.~L.~Altshuler, D.~Khmel'nitzkii, A.~I.~Larkin, and P.~A.~Lee,
Phys.~Rev.~B \textbf{22}, 5142 (1980).
% ``Magnetoresistance and Hall effect in a disordered two-dimensional electron gas''
% http://dx.doi.org/10.1103/PhysRevB.22.5142

\bibitem{Bergmann}
G.~Bergmann,
Phys.~Rep.~\textbf{107}, 1 (1984).
% ``Weak localization in thin films: a time-of-flight experiment with conduction electrons''
% http://dx.doi.org/10.1016/0370-1573(84)90103-0

\bibitem{Montambaux}
E. Akkermans and G. Montambaux,
{ \it Mesoscopic Physics of Electrons and Photons},
(Cambridge University Press,2007).

\bibitem{Hikami-Larkin-Nagaoka}
S.~Hikami, A.~I.~Larkin, and Y.~Nagaoka,
Prog.~Theor.~Phys.~\textbf{63}, 707 (1980).
% ``Spin-Orbit Interaction and Magnetoresistance in the Two Dimensional Random System''
% http://dx.doi.org/10.1143/PTP.63.707

\bibitem{Hikami}
S.~Hikami,
Phys.~Rev.~B \textbf{24}, 2671 (1981).
% ``Anderson localization in a nonlinear-sigma-model representation''
% http://dx.doi.org/10.1103/PhysRevB.24.2671

\bibitem{ILP}
S.~V.~Iordanskii, Yu.~B.~Lyanda-Geller, and G.~E.~Pikus,
JETP Lett.~\textbf{60}, 206 (1994);
% ``Weak localization in quantum wells with spin-orbit interaction''
% http://www.jetpletters.ac.ru/ps/1323/article_20010.shtml
W.~Knap \textit{et al.},
Phys.~Rev.~B \textbf{53}, 3912 (1996).
% ``Weak antilocalization and spin precession in quantum wells''
% http://dx.doi.org/10.1103/PhysRevB.53.3912

\bibitem{Koga_2002_2}
T.~Koga, J.~Nitta, T.~Akazaki, and H.~Takayanagi,
Phys.~Rev.~Lett.~\textbf{89}, 046801 (2002).
% ``Rashba Spin-Orbit Coupling Probed by the Weak Antilocalization Analysis in InAlAs/InGaAs/InAlAs Quantum Wells as a Function of Quantum Well Asymmetry''
% http://dx.doi.org/10.1103/PhysRevLett.89.046801

\bibitem{Olshanetsky_2010}
E.~B.~Olshanetsky \textit{et al.},
JETP Lett.~\textbf{91}, 347 (2010).
% ``Weak antilocalization in HgTe quantum wells near a topological transition''
% http://dx.doi.org/10.1134/S0021364010070052

\bibitem{Chen_2010}
J.~Chen \textit{et al.},
Phys.~Rev.~Lett.~\textbf{105}, 176602 (2010).
% ``Gate-Voltage Control of Chemical Potential and Weak Antilocalization in Bi2Se3''
% http://dx.doi.org/10.1103/PhysRevLett.105.176602

\bibitem{He_2011}
H.-T.~He \textit{et al.},
Phys.~Rev.~Lett.~\textbf{106}, 166805 (2011).
% ``Impurity Effect on Weak Antilocalization in the Topological Insulator Bi2Te3''
% http://dx.doi.org/10.1103/PhysRevLett.106.166805

\bibitem{Checklesky_2011}
J.~G.~Checkelsky, Y.~S.~Hor, R.~J.~Cava, and N.~P.~Ong,
Phys.~Rev.~Lett.~\textbf{106}, 196801 (2011).
% ``Bulk Band Gap and Surface State Conduction Observed in Voltage-Tuned Crystals of the Topological Insulator Bi2Se3''
% http://dx.doi.org/10.1103/PhysRevLett.106.196801

\bibitem{Triscone_2010}
A.~D.~Caviglia \textit{et al.},
Phys.~Rev.~Lett.~\textbf{104}, 126803 (2010).
% ``Tunable Rashba Spin-Orbit Interaction at Oxide Interfaces''
% http://dx.doi.org/10.1103/PhysRevLett.104.126803

\bibitem{LAO-STO}
J.~A.~Sulpizio, S.~Ilani, P.~Irvin, and J.~Levy,
Annual Review of Materials Research \textbf{44},  (2014);
% “Nanoscale Phenomena in Oxide Heterostructures.” Annual Review of Materials Research 44, no. 1 (2014): null. doi:10.1146/annurev-matsci-070813-113437.
S.~Stemmer and S.~J.~Allen,
Annual Review of Materials Research \textbf{44},  (2014);
%“Two-Dimensional Electron Gases at Complex Oxide Interfaces.” Annual Review of Materials Research 44, no. 1 (2014): null. doi:10.1146/annurev-matsci-070813-113552.
H. Y. Hwang, Y. Iwasa, M. Kawasaki, B. Keimer, N. Nagaosa, and Y. Tokura,
Nature Materials \textbf{11}, (2012).
% “Emergent Phenomena at Oxide Interfaces.” Nature Materials 11, no. 2 (February 2012): 103–13. doi:10.1038/nmat3223.


\bibitem{Guru}
G.~Khalsa, B.~Lee, and A.~H.~MacDonald,
Phys.~Rev.~B \textbf{88}, 041302 (2013).
% "Theory of t2g Electron-Gas Rashba Interactions."
%  doi:10.1103/PhysRevB.88.041302.

\bibitem{Zhong}
Z.~Zhong, A.~T\'{o}th, and K.~Held,
Phys.~Rev.~B \textbf{87}, 161102 (2013).
% "Theory of Spin-Orbit Coupling at LaAlO3/SrTiO3 Interfaces and SrTiO3 Surfaces."
%   doi:10.1103/PhysRevB.87.161102.

\bibitem{Joshua}
A.~Joshua, S.~Pecker, J.~Ruhman, E.~Altman, and S.~Ilani,
Nat.~Commun.~\textbf{3}, 1129 (2012).
% ``A universal critical density underlying the physics of electrons at the LaAlO3/SrTiO3 interface''
% doi:10.1038/ncomms2116

\bibitem{DiXiao_2012}
D.~Xiao, G.-B.~Liu, W.~Feng, X.~Xu, and Wa.~Yao,
Phys.~Rev.~Lett.~\textbf{108}, 196802 (2012).
% ``Coupled Spin and Valley Physics in Monolayers of MoS2 and Other Group-VI Dichalcogenides''
% http://dx.doi.org/10.1103/PhysRevLett.108.196802

\bibitem{DiXiao_2013}
H.-Z.~Lu, W.~Yao, D.~Xiao, and S.-Q.~Shen,
Phys.~Rev.~Lett.~\textbf{110}, 016806 (2013).
% ``Intervalley Scattering and Localization Behaviors of Spin-Valley Coupled Dirac Fermions''
% http://dx.doi.org/10.1103/PhysRevLett.110.016806

\bibitem{Neal_2013}
A.~T.~Neal, H.~Liu, J.~Gu , and P.~D.Ye,
ACS Nano \textbf{7}, 7077 (2013).
% ``Magneto-transport in MoS2: Phase Coherence, Spin–Orbit Scattering, and the Hall Factor''
% http://dx.doi.org/10.1021/nn402377g

\end{thebibliography}
\end{document}